\documentclass[twocolumn,
superscriptaddress,
prl,
showpacs,
preprintnumbers,
amsmath,amssymb]{revtex4}

\usepackage{graphicx,epsf,epsfig,ulem}
\usepackage{epsfig}
\usepackage{epstopdf}

\usepackage{bm}
\usepackage{subfigure}
\usepackage{color}

\begin{document}


\title{Bulk and sub-surface donor bound excitons in silicon under electric fields}

\author{Rajib Rahman*}
\affiliation{Network for Computational Nanotechnology, Purdue University, West Lafayette, IN 47907, USA}

\author{Jan Verduijn}
\affiliation{Centre for Quantum Computation and Communication Technology, School of Physics, University of New South Wales, Sydney, NSW 2052, Australia}

\author{Yu Wang}
\affiliation{Network for Computational Nanotechnology, Purdue University, West Lafayette, IN 47907, USA}

\author{Chunming Yin}
\affiliation{Centre for Quantum Computation and Communication Technology, School of Physics, University of New South Wales, Sydney, NSW 2052, Australia}

\author{Gabriele De Boo}
\affiliation{Centre for Quantum Computation and Communication Technology, School of Physics, University of New South Wales, Sydney, NSW 2052, Australia}

\author{Gerhard Klimeck}
\affiliation{Network for Computational Nanotechnology, Purdue University, West Lafayette, IN 47907, USA}


\author{Sven Rogge}
\affiliation{Centre for Quantum Computation and Communication Technology, School of Physics, University of New South Wales, Sydney, NSW 2052, Australia}

\date{\today}

\begin{abstract}
The electronic structure of the three-particle donor bound exciton (D$^0$X) in silicon is computed using a large-scale atomic orbital tight-binding method within the Hartree approximation. The calculations yield a transition energy close to the experimentally measured value of 1150 meV in bulk, and show how the transition energy and transition probability can change with applied fields and proximity to surfaces, mimicking the conditions of realistic devices. The spin-resolved transition energy from a neutral donor state (D$^0$) to D$^0$X depends on the three-particle Coulomb energy, and the interface and electric field induced hyperfine splitting and heavy-hole-light-hole splitting. Although the Coulomb energy decreases as a result of Stark shift, the spatial separation of the electron and hole wavefunctions by the field also reduces the transition dipole. A bulk-like D$^0$X dissociates abruptly at a modest electric field, while a D$^0$X at a donor close to an interface undergoes a gradual ionization process. Our calculations take into account the full bandstructure of silicon and the full energy spectrum of the donor including spin directly in the atomic orbital basis and treat the three-particle Coulomb interaction self-consistently to provide quantitative guidance to experiments aiming to realize hybrid opto-electric techniques for addressing donor qubits.    

\end{abstract}


\maketitle 

Nuclear spins in semiconductors are promising candidates for solid-state qubits with the advantage of exceptionally long spin coherence times and the potential to be integrated into the existing semiconductor device technology \cite{Kane, Morello_nqubit, Si29_qubit, Diamond_qubit1}. Recent experiments have shown that donor nuclear spins in enriched silicon can achieve coherence times exceeding 30 minutes even at room temperature due to a semiconductor vacuum-like environment for spins \cite{Thewalt_science2}. However, the relative isolation of the nuclear spins from their environment also makes it difficult to prepare, and readout their spin states. Several recent experiments have utilized optical techniques to initialize and readout the donor nuclear spins in silicon by photoexcitation of a donor bound exciton - a three-particle system with two electrons and a hole bound to the donor \cite{Thewalt_science1, Morton_D0X, Thewalt_PRL1, Thewalt_PRL2, Rogge_Erbium}. While these experiments have so far been performed on bulk donor ensembles, the techniques will offer the best rewards if extended to single donor spin qubits in the already existing design schemes of a silicon quantum computer \cite{Hollenberg_2D, RMP}. Unlike the dominant spin readout techniques that rely on spin-to-charge conversion, optical readout can offer higher fidelities as it is not limited by temperature dependent broadening in the leads. 

\begin{figure}[htbp]
\center\epsfxsize=3.4in\epsfbox{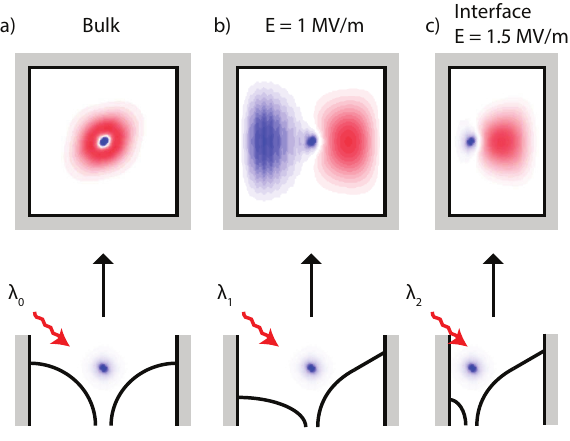}
\caption{ Schematic of the photon induced transition from a Coulomb bound electron in a donor to a donor bound exciton in bulk a) with and b) without an electric field, and c) in a confined nano device with an interface and electric field. The blue and the red color describes the electron and hole probability densities respectively. The bottom panel provides a 1D schematic of the Coulomb potential of the donor core.}

\vspace{-0.5cm}
\label{fi1}
\end{figure}

In this letter, we investigate the electronic structure of the donor bound exciton (D$^0$X) in silicon under the influence of electric fields and interfaces, thus mimicking the environment of realistic devices. This is required since single spin readout in a quantum computer can not be achieved in a bulk environment but requires a device environment. The electronic states of the donor bound exciton are influenced by the complicated bandstructure of silicon including its multiple valley degeneracy in the conduction band (CB), the light and heavy hole valence bands (VB), the Coulomb bound donor states and the three-particle fermionic interactions. We include all these factors in quantitative details using an atomistic tight-binding method over a silicon lattice of one million atoms, and compute the spin-resolved transition energies and transition dipoles of the D$^0$X in a gated nanostructure. For specificity, we refer here only to the most common shallow donor in silicon, phosphorus.  These methods can readily be generalized to the other shallow donors Sb, As and Bi, which have slightly different electronic binding energies, and also different nuclear spins and hyperfine constants.

Three-partcile excitons or trions have also been reported recently in two-dimensional materials such as MoS$_2$ \cite{Trions_NMat, Chernikov_PRL} and have attracted much attention as a test-bed for studying many-body interactions in a semiconductor with both spin and valley degrees of freedom. Previous theoretical studies in bulk D$^0$X in silicon have been performed from effective mass approximations to understand Auger recombination in donors \cite{Auger_paper} and more recently to investigate applications such as quantum Hall charge sensors \cite{QH_charge_sensor}. This work presents atomistic electronic structure calculations of the D$^0$X in a device environment for the first time, crucial for experiments to integrate optical addressing methods in a silicon quantum computer.

  
A neutral P donor in bulk silicon comprises a positively charged donor core that produces a Coulomb potential and binds an electron 45 meV below the conduction band edge (also called the D$^0$ state). If stimulated by a laser pulse of appropriate wavelength, an additional electron and a hole can be generated and bound to the impurity, forming the donor bound exciton (D$^0$X), as shown in Fig. 1a) for a bulk donor, with blue and red representing the electron and hole wavefunction probability densities respectively. The bulk transition energy from a D$^0$ to a D$^0$X state has been measured to be near 1150 meV \cite{Thewalt_science2}. Fig. 1b) shows a schematic of the same system under an applied electric field. Both the D$^0$ and the D$^0$X undergo a Stark shift, however the latter is more strongly perturbed due to the larger size of the wavefunction. The electric field pushes the electrons and holes in opposite directions and thus changes the excitation wavelength from $\lambda_0$ to $\lambda_1$. Fig. 1c) shows the combined effect of an interface and an electric field with excitation wavelength $\lambda_2$. In this case, the donor wave functions are also truncated by the interface. Later, we will show the transition energies, Hartree energy, wavefunctions and transition dipole moments for these three cases, as well as the interface and E-field induced hyperfine and hole splitting that can affect the spin-resolved transition energies.

\begin{figure}[htbp]
\center\epsfxsize=3.4in\epsfbox{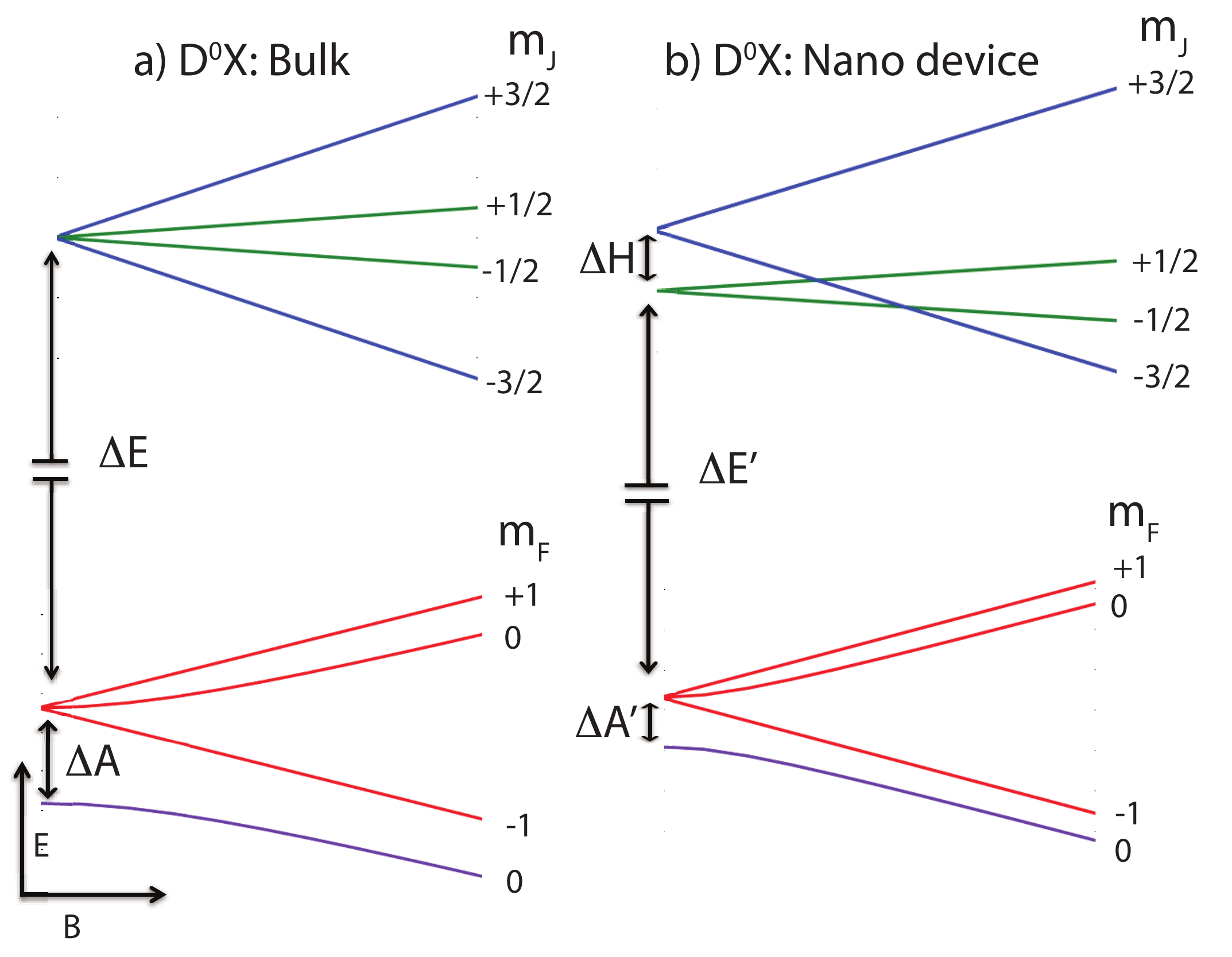}
\caption{ Energy levels and spin structure of the D$^0$X in a magnetic ($B$) field  a) in bulk and b) in a nanostructure. $\Delta A$ ($\Delta A'$) and $\Delta H$ denote the hyperfine and heavy hole-light hole splitting respectively, while $\Delta E$ ($\Delta E'$) is the formation energy at $B=0$.}
\vspace{-0.5cm}
\label{fi2}
\end{figure}

In Fig. 2, a schematic of the energy levels of the D$^0$ and D$^0$X are shown for a bulk-like exciton and an exciton in a nanoscale device. In the D$^0$ state, the hyperfine interaction between the nuclear and the electronic spins produces a singlet and a triplet splitting $\Delta A$. The D$^0$X has two electrons spin paired in a singlet with $S=0$, and as a result, the spin splitting in a B-field is governed by the hole 3/2 and 1/2 total angular momentum states, which arise from the heavy hole (HH) and the light hole (LH) bands. If the donor is located close to an interface or subjected to an electrical field, then the transition energies of the D$^0$X are strongly modified, as shown in Fig. 2b). First, the presence of an interface or electric field changes the localization of all the confined particles in both D$^0$ and D$^0$X. As the wave functions shift and deform in space, the particle-particle interactions also change. We have computed this transition energy from self-consistent Hartree method with tight-binding, and denote it as $\Delta E'$. Second, the electric field or the interface also splits the 3/2 and 1/2 states of the hole bound to the D$^0$X, which we denote as $\Delta H$. This splitting is due to the effect of confinement on the HH and LH bands, which have different effective masses. Third, the hyperfine coupling of the D$^0$ state also changes due to the field and depth \cite{Rahman_prl}, and is denoted as $\Delta A'$ in Fig. 2b). Details of how the spin-resolved D$^0$X transitions can be used to initialize and readout the donor nuclear spins using an Auger recombination process can be found in Refs \cite{Thewalt_science2, Thewalt_science1}.

\begin{figure}[htbp]
\center\epsfxsize=3.4in\epsfbox{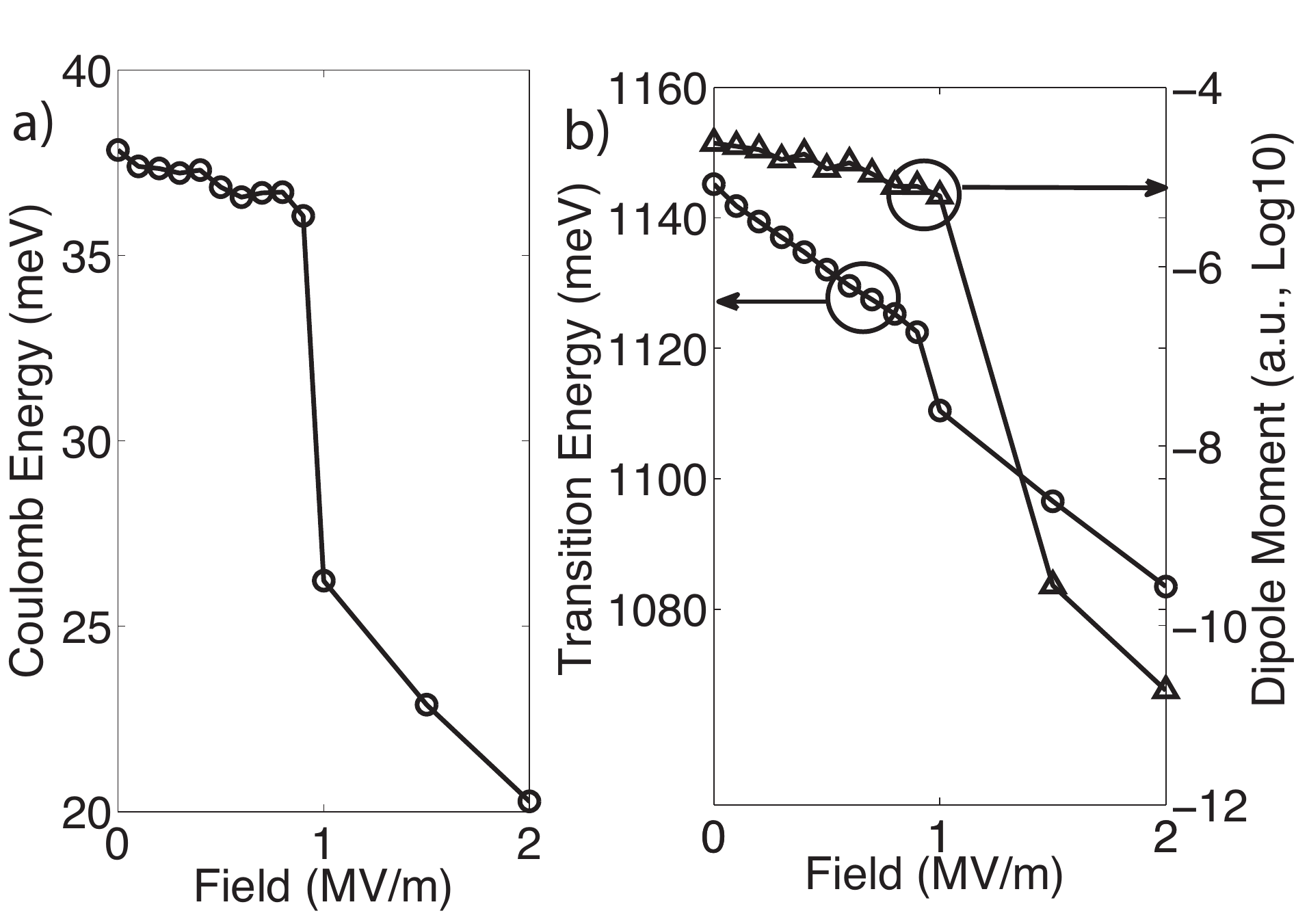}
\caption{ a) Coulomb energy of the three-particle system with E-field. b) Transition energy (left axis) from D$^0$ to D$^0$X for a bulk-like system as a function of an applied E-field. Transition dipole moment (right axis) from D$^0$ to D$^0$X with E-field.}
\vspace{-0.5cm}
\label{fi3}
\end{figure}

We employ an sp$^3$d$^5$s* atomic orbital tight-binding method with spin-orbit interaction and nearest neighbor coupling to represent the full band structure of silicon \cite{Rahman_prl, Klimeck_ted, Shaikh, Rogge_NPhys, Salfi_NMat, Rahman_ce}. More details of the methods can be found in the supplementary materials. First, we investigate the Stark effect of the D$^0$X when the donor is in bulk silicon. Fig. 3 shows the Coulomb energy, transition energy and transition dipole moment of the three-particle D$^0$X for a bulk-like case as a function of an electric field. In the simulation, a bulk system is represented by a 30 nm $\times$ 30 nm $\times$ 30 nm of silicon lattice bounded by hydrogen passivated hard-walls on all sides. The donor is placed at the center of the box, and the simulation domain size is chosen such that the electronic and hole wavefunctions at zero E-field are not affected by the boundaries of the box. This is reasonable as the Bohr radii of the D$^0$ and D$^-$ states are in the range of 1-4 nm. 

The Coulomb energy of the D$^0$X comprises of the repulsive energy between the two electrons and the attractive energy between the hole and each of the electrons. Fig. 3a) shows this three-particle Coulomb energy with an E-field, computed from eq 3 of the Supplementary Materials (SM). At low fields, the Coulomb energy lies between 35 and 40 meVs, and decreases slightly with the field. It is known that the Coulomb energy in the two-electron D$^-$ state is about 43 meV at zero E-field \cite{Rahman_ce}. Therefore, the attractive energy of the hole in D$^0$X reduces the total Coulomb energy to sub 40 meV. As the E-field is increased, the electron and hole wavefunctions are spatially separated steadily, and the Coulomb energy decreases. An abrupt drop in the Coulomb energy is observed in Fig. 3a) at about F=1 MV/m as the hole and one of the electrons ionize to opposite surfaces of the bounding box. The E-field produces triangular wells at the opposite surfaces (in the CB and VB). When the energy of the bound states in these triangular wells are lower than those of the Coulomb bound states (including mutual screening between the particles), the electrons and holes localize at the surfaces. As a result, the large spatial separation between the three particles at ionization reduces their net electrostatic interaction.      

The transition energy required to excite a D$^0$ into a D$^0$X state is computed from the difference of the total energies of the D$^0$X and D$^0$ as given by eq 6 of the SM, and is shown in Fig 3b) on the left vertical axis. Using the low temperature bandgap of 1.17 eV in silicon, the transition energy at zero field (F=0) is 1145 meV in close agreement with the experimental measurements \cite{Thewalt_science1}. It is to be noted that no fitting parameters are used in the D$^0$X simulations. As the E-field is increased, the transition energy drops gradually untill an abrupt drop is witnessed at F=1 MV/m due to field ionization. Overall the transition energy decreases with E-field due to both the reduced Coulomb energy and the reduced energy difference between the single particle orbital energies in the CB and in the VB. 

A reduced transition energy with E-field, however, does not mean that the D$^0$ to D$^0$X transition becomes more likely. The transition probability is governed by the dipole moment between the wavefunctions which depends strongly on their spatial overlaps. Using eq 7 of the SM, we evaluated the transition dipole moment of the D$^0$X in Fig. 3b) on the right vertical axis. It is shown that the dipole moment is large at low E-fields and close to the F=0 value, but decreases rapidly by orders of magnitude near F=1 MV/m. As the hole and one of the electrons are ionized towards opposite surfaces, the spatial overalp of the wavefunctions of the three particles diminish abruptly, making the D$^0$ to D$^0$X excitation unlikely. 

\begin{figure}[htbp]
\center\epsfxsize=3.4in\epsfbox{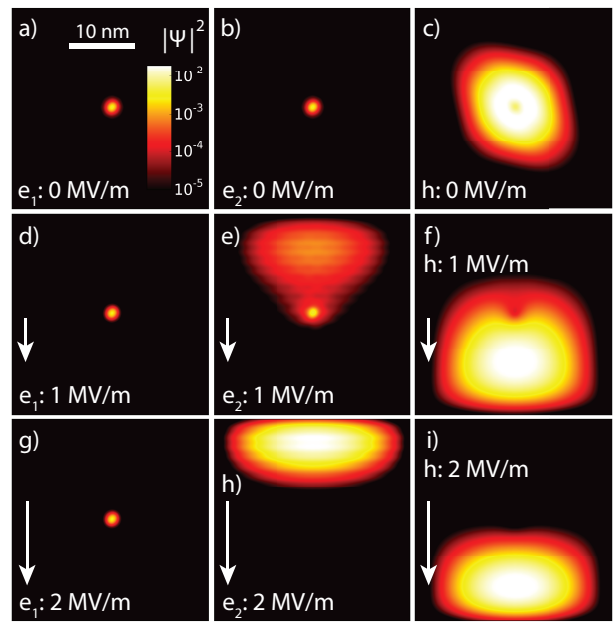}
\caption{ Wavefunction probability density ($| \psi |^2$) of the two electrons (column 1 and 2) and the hole (column 3) plotted separately for a bulk-like donor. Each row shows $| \psi |^2$ for a specific E-field. a-c) are for F=0, d-f) for F=1 MV/m, and g-i) for F=2 MV/m.}
\vspace{-0.5cm}
\label{fi4}
\end{figure}

Fig. 4 shows the spatial probability densities of the electron and hole wavefunctions with an E-field applied along the vertical direction.The first row depicts the probability distributions (a-c) without an E-field, showing the electrons to be tightly bound to the donor as in the D$^-$ state \cite{Rahman_ce}. Since the hole experiences a weak negative potential of the two electrons, its probability distribution spreads out over a larger region, but is still far from the edges of the bounding box. Hence, the use of 30 nm $\times$ 30 nm $\times$ 30 nm box to represent the bulk-case is justified. The hole wavefunction assumes a more p-like structure as shown by the lower density at the core (green) and a larger density rim (red) surrounding it. As a vertical E-field of 1 MV/m is applied in the second row, one of the electrons remains donor bound (d), while the other electron is pulled opposite the E-field towards the top surface (e). The hole, on the other hand, is pushed towards the back surface in a direction along the E-field (f). Hence, the overlap between the three particles decreases. However, both the electrons and the holes still have some density at the core, showing that the ionization at this field is only partial. The 3rd row shows the complete ionization case at a large E-field of 2 MV/m, in which the electron (h) and hole wavefunctions (i) are fully localized near the surface. At even larger biases, the donor bound electron (g) will eventually also ionize to the surface, but this regime is not pertinent to this work.         

\begin{figure}[htbp]
\center\epsfxsize=3.4in\epsfbox{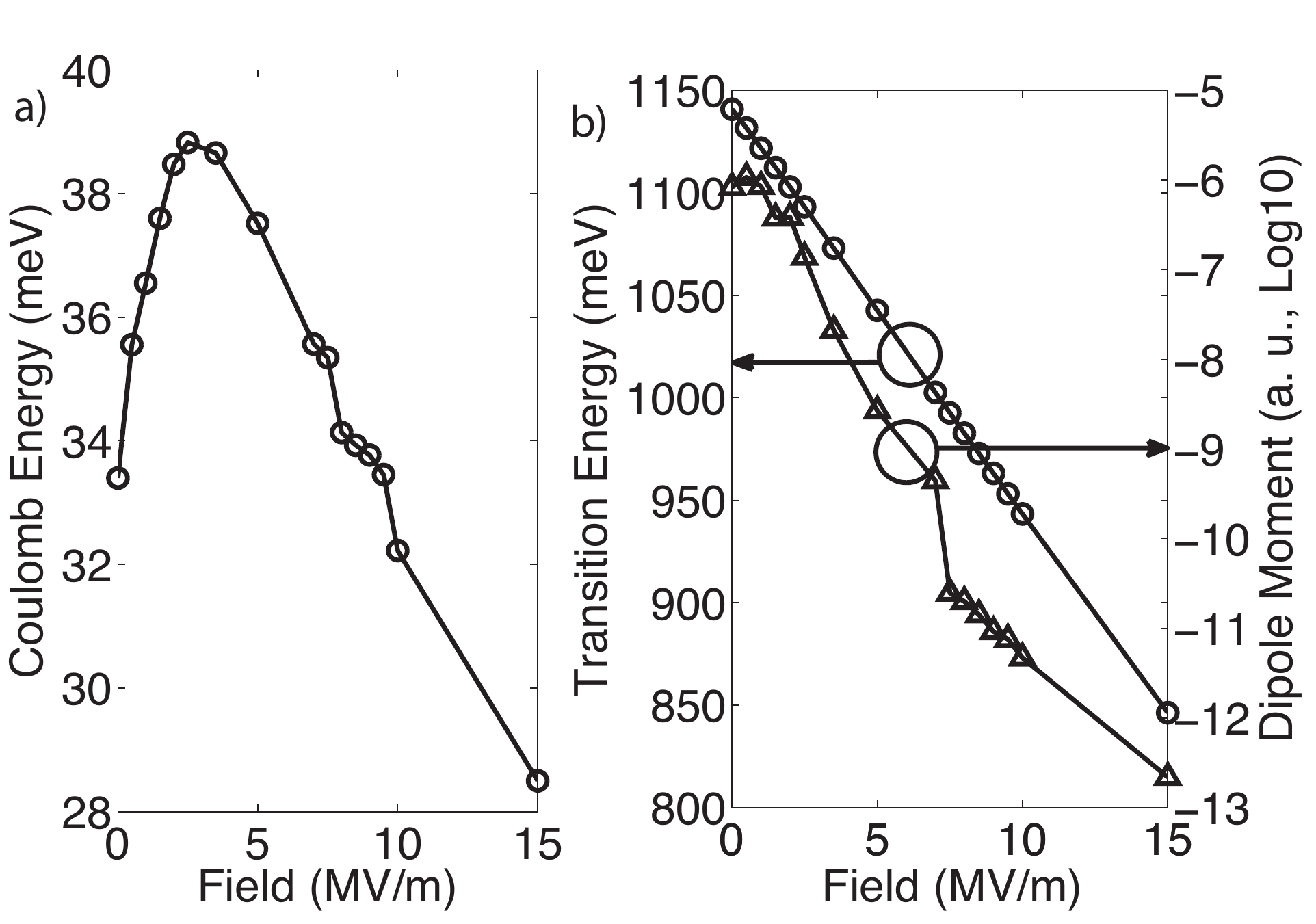}
\caption{ a) Coulomb energy of D$^0$X for a donor 4.2 nm from an interface as a function of  E-field. b) Transition energy (left) and dipole moment (right) for the same.}
\vspace{-0.5cm}
\label{fi5}
\end{figure}

Next, we investigate the Stark effect of the D$^0$X when the donor is closer to a surface, which is more representative of a realistic device. It was shown in prior works \cite{Rahman_orbital_stark} that a single electron bound to a donor in silicon can undergo strong orbital Stark shift if the donor depth D from the surface is small (typically $<$ 6 nm). In this case, a large E-field can be applied without completely ionizing the donor electron. At these relatively large E-fields, the donor orbitals hybridize with the confined states of the triangular well at the surface, resulting in a gradual (non-abrupt) ionization process. 

Fig. 5a) shows the three-particle Coulomb energy for the D$^0$X when the donor is at 4.2 nm depth from the surface. The Coulomb energy is still in the sub 40 meV range, but decreases slightly below the bulk value at F=0. The high potential barrier at the surface not only truncates part of the wavefunction but also pushes the asymmetric charge densities a bit off the donor center with the electron wavefunctions slightly shifting deeper into the lattice and the hole wavefunction moving closer to the surface. The Coulomb energy is maximum when the electron-hole charge densities are centered around the donor core, but decreases when the electron-hole wavefunctions are displaced from the donor core. A small applied E-field acts against the surface potential and pulls the electron-hole densities towards the donor core. Hence, the Coulomb energy is seen to increase at low E-fields and reaches a maximum of 39 meV. A larger applied field now pushes the electrons towards the surface and holes away from the surface (acting opposite the E-field due to the surface potential), and decreases the Coulomb energy monotonically. Unlike the bulk case, the ionization process is gradual (non-abrupt), and the Coulomb energy curve appears smooth even when the hole and one of the electrons ioinize at about 10 MV/m. Fig. 5b) also shows the transition energy and dipole moments as before, however, these curves are also relatively smoother as the larger E-fields cause a gradual ionization of the electrons and holes. 

The ionizing field for the D$^0$X is depth dependent. Ignoring the donor potential and the inter-particle interactions, the potential drop between the donor location and the surface for electrons and holes is $qF \rho_{e,h}$, where, $\rho_e$ is equal to the donor depth, $D$, and $\rho_h$ equals the distance to the opposite surface, $W-D$, in which $W$ is the thickness of the channel. Hence, even for a donor close to an interface (small $D$) the D$^0$X disassociates already at moderate field strengths by removal of the hole if $W$ is large compared to $D$. This is qualitatively different from field ionization of the D$^0$ and D$^-$ states \cite{Rahman_orbital_stark, Rahman_ce}.



\begin{figure}[htbp]
\center\epsfxsize=3.4in\epsfbox{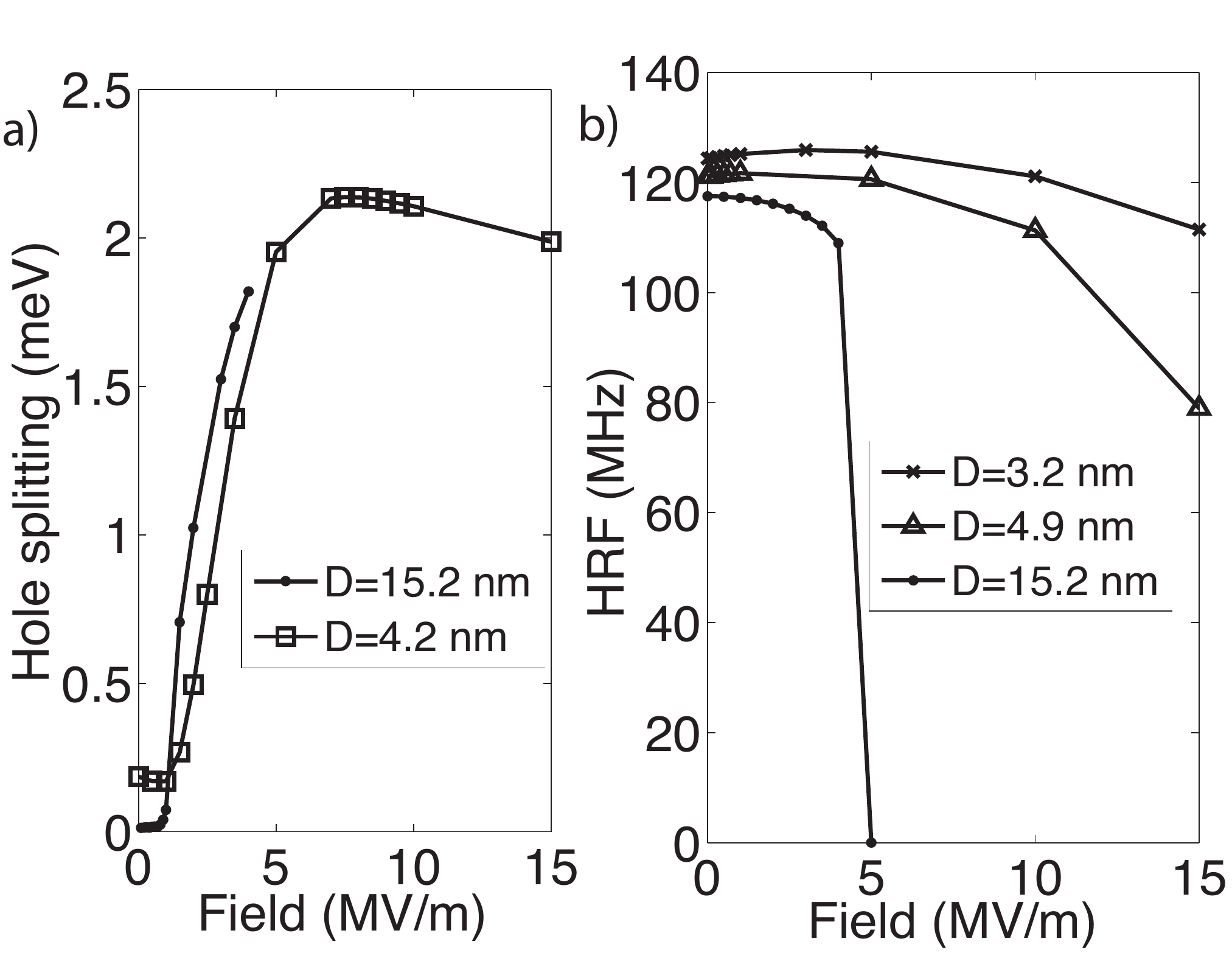}
\caption{ a) D$^0$X hole splitting with E-fields for a bulk and a sub-surface dopant. b) Hyperfine resonance frequencies (HRF) of a P donor below a depth D from the silicon surface with applied electric fields. }
\vspace{-0.5cm}
\label{fi6}
\end{figure}

In Fig. 6, we investigate the other two energy gaps, $\Delta H$ and $\Delta A$, that influence the D$^0$ to D$^0$X transition energy, as shown in Fig. 2b). The proximity to an interface and/or the presence of an applied field changes both the LH and HH splitting $\Delta H$, and the hyperfine resonance energy $\Delta A$. Fig. 6a) shows that $\Delta H$ increases with E-field both for bulk and interfacial donors, as the $m_J= \pm 3/2$ and $m_J= \pm 1/2$ states are affected differently, which is consistent with recent studies of Boron acceptors in silicon \cite{Jan_APL}. Fig. 6b) shows the variation of the hyperfine coupling with E-field for donors at various depths. The shift in the hyperfine coupling from a bulk donor can be directly evaluated using the amplitude of the D$^0$ wavefunction at the donor site, as done in Ref \cite{Rahman_prl}. $A$ for bulk-like donors (D=15.2 nm) varies from the bulk value of 117 MHz at small fields before abruptly diminishing to 0 as the donor ionizes. In contrast, $A$ for donors nearer to an interface can assume a continuous range of values (117 to 0 MHz) depending on the field as the ionization process is gradual. Also, much higher fields in the range of 10-40 MV/m can be applied to donors less than 5 nm deep without completely ionizing the D$^0$ electron. 


There are two factors which we have ignored here, as they lie beyond the scope of this work. First, we have neglected exchange and correlation effects, and treated the problem within the Hartree approximation for computational feasibility. There is a 5 meV difference between the calculated and measured bulk transition energy \cite{Thewalt_science1}, as shown earlier, which could be due to this approximation. More justification of this approximation is provided in the SM. Second, some residual strain has been reported recently in nano devices close to the oxide-silicon interface \cite{Thorbeck}. Although we have not included the effect of local strain in the calculations as it may vary from device to device, qualititatively the effect of strain is similar to that of the electric field. The strain may cause valley-repopulation of the donor, and reduce the weight of the electronic wavefunction in the central-cell. As a result, strain will reduce both the Coulomb energy $\Delta E$ and the hyperfine coupling $\Delta A$. In addition, strain removes the light and heavy hole degeneracy, and enhances $\Delta H$, which has been shown in Ref \cite{Morton_D0X}.   


We have shown that electric fields and interfaces can affect the transition energies and dipole moments of donor bound excitons in nanodevices quite siginificantly, and optical addressing techniques for donor qubits need to be aware of these changes to be successful. We have also performed the first atomistic electronic structure calculation of the donor bound exciton in silicon taking into account the full bandstructure and full donor energy spectrum using spin-resolved atomic orbitals of over a million atoms, and have obtained the three-particle Coulomb energy, hyperfine and hole splittings, all of which contribute to the D$^0$X transition energy. A detailed understanding of the range of excitation frequencies possible in nanoscale devices and the environmental conditions needed for a stable D$^0$X helps to extend the optical addressing methods from bulk ensembles to single donor qubits, and may also enable hybrid opto-electric addressing of nuclear spins in silicon with improved fidelity and possible higher temperature operation.

\begin{acknowledgments}
This research was conducted by the Australian Research Council Centre of Excellence for Quantum Computation and Communication Technology (Project number CE110001027). Financial support from the the U.S. National Security Agency (NSA) and the Army Research Office (ARO) under Contract No. W911NF-04-1-0290 is also acknowledged. C. Y. acknowledges an Australian Research Council Discovery Early Career Researcher Award (DE150100791). S.R. acknowledges an Australian Research Council Centre Future Fellowship (FT100100589). NEMO 3D was initially developed at JPL, Caltech under a contract with the NASA. NSF funded NCN/nanoHUB.org computational resources were used. We acknowledge discussions and feedback from Mike Thewalt.
\end{acknowledgments}

Electronic address: rrahman@purdue.edu


\section{Supplementary Information: Method Details}

The Hamiltonian of the three-particle D$^0$X is given by,

\begin{equation}  
H=H_{0,e_1}+H_{0,e_2}+H_{0,h} + H_{e_1,e_2} +H_{e_1,h}+H_{e_2,h} \label{vgl1}
\end{equation}
\noindent
where the first three terms represent the single particle Hamiltonian of the two electrons ($e_1$, $e_2$), and the hole $h$, and the last three terms are the pairwise Coulomb interaction between the three particles. Each of the $H_{0, n}$, for $n \in \{ {e_1, e_2, h} \}$ includes the Hamiltonian of the silicon lattice $H_{Si}$, the potential energy of the donor atom $V_D$, and the potential energy of the applied electric field $F$, and is given by,

\begin{equation}  
H_{0,n}=H_{Si}+V_D-e\vec{F} \cdot \vec{r} \label{vgl2}
\end{equation}   

In the atomistic tight-binding formalism employed here, $H_{Si}$ is represented in a basis of 20 atomic orbitals (sp$^3$d$^5$s*) per atom including spin in the basis and nearest neighbor interactions \cite{Klimeck_ted}. For a bulk silicon unit cell, the model reproduces the full bandstructure of the host material \cite{Boykin}. Surfaces are treated as dangling bond passivated following the model shown in Ref \cite{Passivation_Lee}. The donor potential energy is represented as a Coulomb potential with an onsite cut-off term and an orbital based central-cell correction, and reproduces the full spectrum of donor bound states including a ground state at 45.6 meV below the conduction band \cite{Rahman_prl, Shaikh}. For the solution of a bulk single donor in silicon, a box size of 30 nm $\times$ 30 nm $\times$ 30 nm was used such that the donor wavefunctions are not affected by the interfaces of the box \cite{Rahman_iedm}. The model has been used in a variety of earlier works to reproduce various experimental data with very good accuracy \cite{Rahman_prl, Rogge_NPhys, Salfi_NMat, Hsueh}. The resulting Hamiltonian is solved by a parallel block lanczos eigen solver to obtain a set of states close to the conduction and valence band edges \cite{Lee_jce}.     


An exact solution of the three-particle system involves diagonalizing eq. 1 in the basis of many Slater Determinants comprising of the single particle electron and hole states until convergence is achieved with the so called Configuration Interaction (CI) approach. However, such an approach is computationally intractable for a Hamiltonian of over one million atoms, as Coulomb and exchange integrals between the three particles need to be computed with a very large number wave functions to represent the spatial spread and symmetries of the electrons and holes. Hence, we employ a mean-field method based on the self-consistency between the effective single particle Hamiltonian of each particle and the Poisson equation. In this approach, the particles are each represented with a separate Hamiltonian taking into account the mutual interactions, as described below,     

\begin{equation} 
H_{e_1}=H_{0,e_1}+V_{e_2}+V_h \label{vgl3}
\end{equation}

\begin{equation}
H_{e_2}=H_{0,e_2}+V_{e_1}+V_h \\
\end{equation}

\begin{equation}
H_{h}=H_{0,h}+V_{e_1}+V_{e_2}
\end{equation}


Here, each particle is subjected not only to the host and donor potential along with the applied field ($H_{0,n})$, but also to the potential due to the other two particles ($V_n$). Finding an approximate solution can be done by initially solving the three Hamiltonians with all $V_{n} = 0$ to obtain the respective initial wave functions $\Psi_{0,e_1}$, $\Psi_{0,e_2}$, and $\Psi_{0,h}$, and corresponding energies $E_{0,e_1}$ , $E_{0,e_2}$ and $E_{0,h}$. Using these wavefunctions, the charge densities $n(r)$ and the potential $V_n$ can then be computed. Using these new $V_n$ terms, the Hamiltonians are solved again, and the process repeated till the energies and the wavefunctions converge. A similar technique had been used to compute the D$^-$ (two-electron) state of a P donor in close agreement with experimental measurements \cite{Rahman_ce}, and more recently to obtain charging energies of multi-electron donor clusters \cite{Bent_DQD}.

If the final ground state energies of the Hamiltonians shown in eq 3-5 are given by $E_{f, e1}$ , $E_{f, e2}$ and $E_{f,h}$ with $f$ denoting the final iteration ($\Psi_{f,e_1}$, $\Psi_{f,e_2}$, and $\Psi_{f,h}$ being the corresponding wavefunctions), then the total three-particle interaction energy $E_C$ is given by,

\begin{equation}  
E_{C}=\frac{1}{2}[E_{f,e1}+E_{f,e2}+E_{f,h}-E_{0,e1}-E_{0,e2}-E_{0,h}] \label{vgl4}
\end{equation}

Also, the total energy $E_T$ is given by,

\begin{equation}  
E_{T}=E_{0,e1}+E_{0,e2}+E_{0,h}+E_{C} \label{vgl5}
\end{equation}    


The transition energy from the D$^0$ state to a D$^0$X state depends on the difference of the total energy of the two systems, and is given by, 

\begin{equation}
\Delta E=E_T(D^0X)-E_T(D^0) 
\end{equation} 

\noindent
where $E_T(D^0X)$ and $E_T(D^0)$ are the total energies of D$^0$X and D$^0$ respectively. In our calculations, the total energies are with respect to the VB edge at F=0 MV/m (i.e. the 0 eV reference point). If $E_D$ is the donor binding energy relative to the CB edge, corresponding to -45.6 meV for a bulk donor at F=0 MV/m, then $E_T(D^0)=E_g+E_D$, where $E_g=1.17$ eV is the low temperature (4K) bandgap of silicon. $E_D$ changes with donor depth and electric field, as shown earlier \cite{Rahman_orbital_stark}. Hence, the transition energy $\Delta E$ is expected to vary with donor depth and applied electric field, as both $E_T(D^0X)$ and $E_T(D^0)$ change. 


The transition dipole moments describe a photon induced transition from the D$^0$ state to a D$^0$X state, and is given by the dipole moment between the electron and the hole wavefunctions,
 
\begin{eqnarray}
W_{D^0->D^0X} &=&   | \langle \Psi_{0,e_{1,2}}|\vec{r} \cdot \hat{e} |\Psi_{f,e_1}\rangle \langle \Psi_{f,h}|\vec{r}|\Psi_{f,e_2}\rangle \nonumber \\
&& + \langle \Psi_{0,e_{1,2}}|\vec{r} \cdot \hat{e}|\Psi_{f,e_2}\rangle \langle \Psi_{f,h}|\vec{r}|\Psi_{f,e_1}\rangle |^2 \label{vgl10}
\end{eqnarray} 

\noindent
where $\hat{e}$ is the polarization direction of the photon and $\vec{r}$ is the dipole operator. The first term represents a transition in which the D$^0$ electron (denoted by $\Psi_{0, e_1}$ or $\Psi_{0, e_2}$, which are the same) forms electron 1 of D$^0$X, while electron 2-hole transition emerge based on the dipole moment. The second term represents the other pathway in which the D$^0$ electron forms electron 2 of D$^0$X, and electron 1-hole transition results due to the photon. If one of the electrons and the hole are now spatially separated by the E-field, $\langle \Psi_{f,h}|\vec{r}|\Psi_{f,e_2}$ or $\langle \Psi_{f,h}|\vec{r}|\Psi_{f,e_1}\rangle$ decrease as a result of reduced wavefunction overlap. Eq 9 therefore provides a measure of the transition probability from D$^0$ to D$^0$X.

An exact electronic structure calculation of D$^0$X involves a three-interacting particle solution of the system Hamiltonian taking into account description of multiple conduction band valleys and features of multiple valence bands. While simplified effective mass envelope function solutions may be possible, it is difficult to incorporate all the band structure effects, such as valley-orbit  and spin-orbit interactions, multi-valley and multi-band descriptions under a unified framework. Furthermore, the hyperfine structure and the spin structure needs to be computed, in addition to the effects of fields and interfaces. It is also cumbersome to perform a density functional calculation, as the wavefunctions of the holes and electrons typically spread out over 30 nm $\times$  30 nm $\times$  30 nm of silicon lattice (1.4 million atoms). The atomistic tight-binding technique employed \cite{Rahman_prl, Klimeck_ted} in this work provides a suitable compromise between computational rigor and complexity, as multi-million atom systems can be simulated at ease including the full band-structure of the host material. Particle-particle interactions can be also incorporated in a computationally tractable manner through a self-consistent mean-field description \cite{Rahman_ce} without using the more exact CI method \cite{Rahman_dqd}. 

The only compromise we make here is that the mean field approach does not capture exchange and correlation effects. However, the singlet electronic state of the D$^0$X can be described well without considering exchange and spin correlations, as shown by Hartree calculations of the D$^-$ state of the donor, which reproduces the charging energy within 1-2 meVs. The two electron state of the donor has higher order orbital correlations since excited orbital states beyond 1s are needed to describe the spatially extent D$^-$ wavefunctions. However, our self-consistent Hartree method iteratively updates Hamiltonian till the mean-field Coulomb interation converges. Hence, the orbital extent is well-captured by this method \cite{Martin_single_atom}. Furthermore, the hole is distinguishable from the electron pairs by its charge, and the indistinguishable particle CI method does not need to be used in this case.

Electronic address: rrahman@purdue.edu


\end{document}